\documentclass[pmlr]{jmlr}

\usepackage{longtable}

\usepackage{booktabs}
\usepackage[load-configurations=version-1]{siunitx} 

\usepackage{ccaption}

\theorembodyfont{\upshape}
\theoremheaderfont{\scshape}
\theorempostheader{:}
\theoremsep{\newline}

\jmlrvolume{85}
\jmlryear{2018}
\jmlrworkshop{Machine Learning for Healthcare}

\title[High-Throughput Machine Learning from Electronic Health Records]{High-Throughput Machine Learning from Electronic Health Records}

\author{\Name{Ross Kleiman} \Email{rkleiman@cs.wisc.edu} 
       \addr Department of Computer Sciences\\
       University of Wisconsin - Madison\\
       Madison, WI, USA
       \AND
       \Name{Paul Bennett}
       \addr Department of Computer Sciences\\
       University of Wisconsin - Madison
       \AND
	   \Name{Peggy Peissig}
       \addr Marshfield Clinic Research Institute\\
       Marshfield Clinic\\
       Marshfield, WI, USA
	   \AND
	   \Name{Richard Berg}
       \addr Marshfield Clinic Research Institute\\
       Marshfield Clinic
	   \AND
	   \Name{Zhaobin Kuang}
       \addr Department of Computer Science\\
       Stanford University\\
	   Palo Alto, CA, USA
	   \AND
	   \Name{Scott Hebbring}
       \addr Marshfield Clinic Research Institute\\
       Marshfield Clinic
	   \AND
	   \Name{Michael Caldwell}
       \addr Marshfield Clinic Research Institute\\
       Marshfield Clinic
	   \AND
       \Name{David Page}
       \addr Department of Biostatistics and Medical Informatics\\
	   Department of Computer Sciences\\
       University of Wisconsin - Madison} 

\begin{document}

\maketitle

\begin{abstract}
	The widespread digitization of patient data via electronic health records (EHRs) has created an unprecedented opportunity to use machine learning algorithms to better predict disease risk at the patient level. 
	Although predictive models have previously been constructed for a few important diseases, such as breast cancer and myocardial infarction, we currently know very little about how accurately the risk for most diseases or events can be predicted, and how far in advance. 
	Machine learning algorithms use training data rather than preprogrammed rules to make predictions and are well suited for the complex task of disease prediction.
	Although there are thousands of conditions and illnesses patients can encounter, no prior research simultaneously predicts risks for thousands of diagnosis codes and thereby establishes a comprehensive patient risk profile. 
	Here we show that such pan-diagnostic prediction is possible with a high level of performance across diagnosis codes. 
	For the tasks of predicting diagnosis risks both 1 and 6 months in advance, we achieve average areas under the receiver operating characteristic curve (AUCs) of 0.803 and 0.758, respectively, across thousands of prediction tasks. 
	Finally, our research contributes a new clinical prediction dataset in which researchers can explore how well a diagnosis can be predicted and what health factors are most useful for prediction. 
	For the first time, we can get a much more complete picture of how well risks for thousands of different diagnosis codes can be predicted.
\end{abstract}

\section{Introduction}
Around the globe, researchers, insurers, health systems, and governments are keenly interested in reducing the costs of healthcare while improving its outcomes. 
A key component of making such an improvement is for both patients and providers to accurately assess patient health risks.
Concurrent with this emphasis on improving outcomes, the use of electronic health records (EHRs) for digitally capturing patient healthcare encounters has risen dramatically \citep{Hsiao2014}. 
Consequently, disease risk prediction tools that can leverage EHR data to assess patient risk have garnered significant attention. 
Concurrently, machine learning algorithms have grown substantially in popularity and there is now a great interest in using machine learning methods to construct disease risk prediction tools from EHR data. 
A natural first question to ask is, ``What is the distribution of accuracies with which all EHR-coded medical events can be predicted?'' This question has not been answered previously, and we do so here for coded diagnoses.

Much of the existing literature related to predicting disease risks via machine learning algorithms focuses on single diseases or small subsets of diseases \citep{Gail1989,Weiss2012,Huang2014,Himes2009}. 
Small-scale learning tasks often allow for manual selection of the features used by the learning algorithm and manual definition of the cases and controls. 
However, the framework to predict a single disease is not sufficient to assess risks for the thousands of possible illnesses a patient can face. 
The current paradigm, of each publication or tool targeting a specific disease, creates a labyrinth of different implementations and data collection assumptions. 
Therefore, to answer the question of how well all diagnoses can be predicted, we cannot simply aggregate the existing models of varying approach and data source but must rather produce a single pipeline capable of to predict all diagnoses at once using the same set of rules, assumptions, and data source. 
Although there has been some work that has predicted multiple patient diagnoses, to the best of our knowledge it has either been done to test representations of patient data \citep{Miotto2016} or in the context of a retrospective phenotyping task \citep{Che2015}, to ``predict'' or specify who really had an event such as an MI, rather than who will have the event in the current year. 
We believe that our work is greater in breadth than the previous research and that we are the first to produce a machine learning pipeline that attempts to learn thousands of unique models predicting the various diagnosis risks a patient may receive in the future.

We hypothesize that it is possible to predict nearly all disease risks simultaneously at a competitive level of performance. 
We call the approach of learning thousands of predictive models at once ``high-throughput machine learning,'' and we call the application of the approach to predicting risks for many diagnosis codes from EHR data ``pan-diagnostic machine learning.'' To test our hypothesis that accurate pan-diagnostic machine learning is possible, we produced a machine learning pipeline that builds a unique model for each diagnosis code in the 9th revision of the International Classification of Disease (ICD-9) (ICD-9 codes were used rather than ICD-10 because (1) we initiated this study before our data source, [REDACTED\_FOR\_REVIEW] Health System (RFRHS), switched to ICD-10 in October, 2015, and (2) all historical data were coded in ICD-9). 
Although ICD-9 codes primarily serve billing purposes and do not always correspond with diseases, they are useful for both their ubiquitous and often standardized presence in healthcare systems and the fundamental role they play in phenotyping patients. 
The pipeline is general enough to be run on ICD-10 codes, with very little modification once sufficient longitudinal data is available.

In this paper, we present a high-throughput machine learning pipeline that learns diagnosis-specific models for risk prediction. 
Starting with only EHR data, the pipeline learns one model per diagnosis code. 
Our modeling approach automatically learns appropriate case-control matching rules, including controlling for pre-existing conditions, for each diagnosis code. 
Furthermore, each model learns rules about which populations of patients are appropriate for the model to be applied to. 
In this way, our pipeline would automatically learn that a model predicting 10-year Alzheimer's disease risk should not be applied to adolescents. 
While our pipeline is designed to require minimal intervention to allow for ease of use, it is also highly customizable and can accommodate prior knowledge such as phenotyping rules for user-specified case-control definitions. 
We showcase such an example with custom rules for identifying individual pregnancies when predicting pregnancy complications. 
In this work we present the results of our pipeline using data from the RFRHS and introduce a new dataset of performance measures and variable importance values for tens of thousands of clinical prediction tasks. 
Moreover, the code for this pipeline is made publicly available so that it may be readily applied to data from other healthcare entities.

\paragraph{Technical Significance}
This work introduces the concept of "high-throughput machine learning" wherein many thousands of modeling tasks are performed from a single dataset.
High-throughput machine learning differs from both multi-class and multi-label machine learning, as there is an inherent sample selection task prior to the learning.
Each prediction task begins with the selection of appropriate cases and control examples based on general user rules that are adapted for each predictive task.
In this way, high-throughput machine learning is a solution for a problem wherein each of the models would have otherwise needed to be hand constructed.

\paragraph{Clinical Relevance}
Secondary use of EHR data has created fertile ground for better predicting patient risk.
Patient risk prediction is already incorporated into clinical workflows through models such as the Framingham model for cardiovascular risk \citep{Dawber1951} and the Gail model for breast cancer risk \citep{Gail1989}.
However, the overwhelming majority of patient outcomes have not been modeled.
In this work we map the landscape of patient risk predictions by constructing models for many thousands of diagnostic codes across several prediction time windows.
This work not only provides insight into how well individual diagnoses can be predicted, but also explores how predictive quality changes across the diagnostic hierarchy and across time.
We believe that this work will both provide a baseline for how well diagnoses can be predicted as well as provide a valuable research dataset in the form of the feature importances of all of our models.

\section{Cohort}

Here we describe the cohort used in our high-throughput machine learning pipeline.

\subsection{Cohort Selection} 
Our dataset consisted of the demographics, diagnoses, labs, procedures, and vitals of 1.5 million patients who received care at the RFRHS. 
Prior to sharing the data, the RFRHS fully de-identified all data by not only removing any protected health information (PHI) but also by de-identifying codes and values. 
All names for ICD-9 codes, procedures, lab tests, and vital measurements were de-identified, and a separate mapping file was produced. 
Additionally, the values for lab and vital measurements were first mapped to a reference range and then de-identified. 
The mapping files for reidentifying data were used only for summary statistics and to properly analyze results and produce figures. 
The [HOME\_UNIVERSITY] Institutional Review Board (IRB) deferred to the RFRHS IRB under the Wisconsin IRB consortium and waiver of consent was obtained from the RFRHS IRB.

Patients with insufficient data were removed from the dataset. 
Our original dataset was comprised of some 1.5 million patients of which 1.1 million met our data minimum requirement. 
We removed patients with less than 4 distinct diagnoses recorded over their lifetime or whose medical data did not span multiple entry dates. 
This was done to help ensure patients in the dataset received regular care at RFRHS. 
If a patient did not have both a sex and date of birth recorded, then he/she also was removed from the dataset. 
Our dataset also contained many ICD-9 codes not related to diseases. 
These non-disease codes included ICD-9 procedure codes and codes beginning with the letter 'E' or 'V'. 
E-codes describe the external cause of a disease or injury and V-codes do not pertain to disease or injury and are used for supplementary documentation purposes. 
We did not build predictive models for V-codes, E-codes, or ICD-9 procedure codes, though these data were kept as candidate (input) features in the training data. 
Additionally, we did not build predictive models for two major categories in the ICD-9 hierarchy: ``Perinatal Diseases'' (ICD-9 codes 760-779) as there was insufficient data in the records of newborns to predict disease, and ``Symptoms, Signs, And Ill-Defined Conditions'' (ICD-9 codes 780-799) as these are not diseases and were outside of the scope of this research. 

\subsection{Feature Choices} 
Patient EHR data in a research data warehouse is distributed across many tables in a relational database, with tables for diagnoses, labs, vitals, procedures, demographics, etc. 
Furthermore, each patient's health record can be view as an irregularly-sampled timeline of medical events. 
In order to use random forests, we needed to convert this rich, irregular patient data into a single relational table with one row, or record, per patient, and with patient features arranged by column. 
To accomplish this, all features were divided into multiple counts of events occurring within time ranges. 
The time ranges we used were ``last 1 year'', ``last 3 years'', ``last 5 years'', and ``ever'' \citep{lantz2016machine}. 
These time ranges are relative to a specific date, that is prior to a case patient's first entry of a diagnosis of interest. 
For example, a surgical procedure 6 months prior to a diagnosis we are predicting would add a single count to all four columns for that surgical procedure feature in a given patient's row. 
Lab value event counts within a time range were further subdivided based on outcome (normal, abnormal, high, low etc.).

Different lab tests have varying numbers of associated outcomes; for example, one test's result could be continuous while another could be discrete, such as either normal or abnormal. 
All continuous lab results were first discretized by the RFRHS using reference ranges. 
We then constructed one feature per (test, reference-value) tuple, and each such unique tuple received separate counts for the four time ranges. 
For example, a blood sodium measurement would have separate time range counts for values of normal, high, low, etc. 
The process of discretizing the continuous lab values and counting (test, reference-value) tuples as unique features was also done for vitals (e.g. 
high/low systolic blood pressure).

\section{Methods}

\subsection{Pan-Diagnostic Machine Learning Experiments}

While our system can use a wide variety of possible machine learning algorithms, we employ random forests \citep{Breiman2001}, which form a prediction by taking a majority vote amongst an ensemble of decision trees. 
Random forests are known for their competitive accuracy \citep{Caruana2006} and resilience to high-dimensional data \citep{Breiman2001}; because of their use of decision trees, they can capture some non-linear interactions of multiple variables without the need to predefine interaction terms; because of their ensemble nature, they are more robust against overfitting than are individual decision trees \citep{Breiman2001}. 
Our dataset consisted of the labs, vitals, diagnoses, procedures, and demographic information of 1.1 million patients from the RFRHS which serves patients in the [REDACTED\_FOR\_REVIEW] region. 
Using these data, we set out to predict all disease-specific ICD-9 code risks prior to patient diagnosis. 
The models can be constructed to make predictions for arbitrary user-defined time-periods prior to diagnosis. 
In this work, we explore how predictive performance varies with time by evaluating models constructed to predict initial diagnosis at seven different time windows: 1-month, 6-months, 2-years, 5-years, 10-years, 15-years, and 20-years. 
It is worth noting that predicting health events with a long forecast window is likely to truly be a disease risk prediction, whereas a shorter window such a 1-month could be considered a blend of risk prediction and diagnosis, as the patient may be undiagnosed for a disease they physiologically already have. 
Of course, where the shift from risk prediction to diagnosis occurs is disease-specific. 
Additionally, we wished to explore the relationship between model efficacy and disease type and did so by leveraging the disease categories present in the ICD-9 hierarchy. 
We evaluated model performance by measuring Area Under the Receiver-Operating Characteristic Curve (AUC) via ten-fold cross-validation, a robust form of hold-out testing commonly used for evaluation in machine learning.

A recurring challenge we faced in many steps of the pan-diagnostic machine learning pipeline was the need for algorithms to be general, flexible, and able to automatically account for differences in the disease mechanisms and affected populations of different diagnoses. 
Several diagnosis codes required specialized refinements to our general approach for case-control matching (see Case-control matching and Dynamic definition refinement), to avoid artificially high accuracies for trivial reasons. 
This arose when predicting risks for diseases that are complications of preexisting states of health. 
For example, when predicting risk for a pregnancy complication, we wish to match a case with a control who is not only female, but also pregnant, and specifically at a similar stage of pregnancy. 
Simply choosing each control to be of the same age and sex as the case makes prediction artificially easy, because pregnancy and pregnancy-stage become accurate predictive features. 
To automatically determine if a diagnosis requires a set of preexisting diagnoses we created a novel approach which we call ``dynamic definition refinement'' (DDR) (see Dynamic Definition Refinement). 
For each diagnosis code, DDR learns a set of prerequisite diagnoses that we require both the case and control to have on their record prior to the prediction date. 
For example, DDR would learn a rule that a patient who has a diabetic complication should previously have diabetes on their record, and hence any matched control should also have diabetes on their record.

\subsubsection{Building random forests}

To construct the random forest models from summary table data and to calculate AUCs we used scikit-learn 0.16.1, an open source machine learning library \citep{Pedregosa2012}. 
We used scikit-learn's implementation of the RandomForestClassifier. 
For each model, 500 trees were constructed and 10\% of features were randomly selected as candidates for each split. 
All other model specifications were default and are listed in Table \ref{tab:rf_params} for completeness.

For each ICD-9 code a separate random forest model was trained on a minimum of 1,000 and a maximum of 10,000 randomly selected case-control paired patients sampled from the dataset of 1.1 million patients. 
While we chose a maximum of 10,000 patients due to computational constraints, many of our models certainly would have benefited from the use of additional case patients, or even a greater control to case patient ratio. 
We randomly sampled a maximum of 10,000 case-control paired patients by first randomly selecting a case patient, and then randomly selecting a control patient that met the criteria that follow. 
If a code did not meet the minimum of 1,000 case-control paired patients, it was not modeled. 
We note that while there were 22,396 unique ICD-9 codes present in our dataset, the majority of these were very infrequently used for diagnosis during clinical care either due to redundancy or lack of prevalence of the corresponding disease. 
By requiring a minimum of 1,000 case-control paired patients we built at most 3,586 models for a given truncation window. 
Cases were considered positive for a diagnosis if the ICD-9 code appeared two or more times in the patient's record ('rule of two') \citep{Rasmussen2014}. 
Controls paired to cases must have never had the diagnosis being modeled recorded in their record. 
Cases and controls were also required to have the same sex and no more than a 30-day difference in date of birth. 
Further matching criteria were utilized and are detailed in the Dynamic Definition Refinement (DDR) and Break Point Analysis (BPA) methods sections. 
After matching, we identified a truncation date for each pair of case-control patients based on the date of the first entry of the diagnosis of interest on the case patient's record and the length of the truncation window. 
If day t was when the case patient had their first entry of the diagnosis being predicted, and the window length is w, then any data following day t-w was excluded from the training data. 
This truncation is essential to prevent class label information from leaking into the training data, which would bias resulting estimates of model performance on future patients, specifically making them overly-optimistic. 
Even more aggressive truncation was applied when we tested prediction ability even further in advance, requiring truncation from one year to 20 years prior to the event to be predicted. 
 Before carrying out cross-validation for a given prediction task, unsupervised feature selection was employed to retain only features populated for more than 1\% of patients (without regard to case vs. 
control label). 
In this work, we chose to use cross-validation rather than the out-of-bag (OOB) estimates for our random forest models because OOB error can in some cases provide an incorrect assessment of model performance \citep{Mitchell2011}.

\subsubsection{Case-control matching}

During case-control matching, two patients are considered to have a sufficiently close date of birth if the date of birth of the control patient is within 30 days of that for the corresponding case patient. 
We did not consider case patients or control patients where the patient's date of birth would be after the truncation date, e.g., we would not train on an adolescent's data when building a model to predict 20-years in advance. 
Furthermore, we considered the concept of a last known contact or the most recent data point in a patient's record. 
We require that control patients have a last known contact greater than or equal to the date of diagnosis of the case patient. 
This helps ensure that the control patient had the potential of being diagnosed but was not.

\subsubsection{Dynamic definition refinement (DDR)}

Due to the massive number of models built, it would be infeasible for a human to individually develop a prerequisite diagnosis list for each model. 
Therefore, we developed DDR to perform most this work, which we could then inspect and update. 
The DDR approach involves two stages: an algorithmic stage to identify potential prerequisite diagnoses and a manual stage to correct over- or under-controlled diagnoses. 
While a serious effort was made to choose realistic controls for each disease and only predict appropriate ICD-9 codes, we realize that pan-diagnostic machine learning is not without limitations and some of the models we produced may still be somewhat overly optimistic, influenced in part by limitations of the ICD-9 hierarchy.

For each diagnosis code, $DX_i$, DDR first identifies all diagnosis-positive patients (established via ``rule of two''). 
Among these patients the algorithm considers all ICD-9 codes on their records prior to their entry of $DX_i$. 
Any diagnosis code that occurred in at least 85\% of the case (positive) patients became a candidate prerequisite diagnosis for $DX_i$. 
To prevent controlling for ICD-9 codes that describe healthcare encounters, DDR does not consider a specific subset of very general ICD-9 codes (see Table \ref{tab:icd9_excl}). 
We found that these general (non-diagnosis-specific) algorithmic refinements were largely effective, but a small number of the pregnancy complication codes failed to have pregnancy as a prerequisite diagnosis. 
Therefore, we manually added ICD-9 V22 (Normal Pregnancy) as a prerequisite diagnosis for all pregnancy complication codes (ICD-9 630-679). 
The figure in S2 Fig. 
shows an example of the DDR matrix that is generated across all ICD-9 code pairings.

\subsubsection{Predicting pregnancy-related complications}

Case-control matching for pregnancy complications required additional consideration beyond the standard date of birth, sex, 'rule of two', and DDR-control that all of the other codes received. 
To ensure cases and controls are in similar stages of pregnancy, we require that their pregnancies must have begun within two weeks of one another. 
The beginning and end dates of a pregnancy, which we use to define the 'pregnancy era', were determined via the use of break point analysis using the 'segmented' package for R \citep{Muggeo2003,Muggeo2008}. 
Across all patients with two or more V-22 (Normal Pregnancy) ICD-9 codes, the time gaps between adjacent codes were calculated. 
These times gaps were then used by break point analysis \citep{Kuang2016a} to calculate a threshold for determining if two adjacent codes in a patient's record belonged to the same pregnancy. 
Breakpoint analysis returned a threshold of 84 days, which is the maximum amount of time two adjacent pregnancy codes on a given patient's record can be spaced from one another and still belong to the same pregnancy era. 
This threshold value was used to determine the unique pregnancy eras, and their associated start dates, for each patient with two or more V-22 ICD-9 codes. 
Finally, a case-control pair was matched with the additional constraint that their pregnancy eras began within 14 days of one another. 
This enforces the simple notion that case-control pairs should roughly be in the same stage of pregnancy. 

\subsubsection{Truncation date analysis}

The experimental results presented in Figure \ref{fig:censorKDE} were generated via kernel density estimation (KDE) \citep{Parzen1962} using the function and parameters detailed in Table \ref{tab:kde_params}. 
For these results, we do not show confidence intervals on the density estimates of the AUC distributions as they are very tight. 
The maximum achievable 95\% confidence interval with 500 cases and 500 controls is  AUC $\pm 0.036$, and with 5,000 cases and 5,000 controls is AUC $\pm 0.011$ (see the simulated prospective study analysis section for details on calculating AUC confidence intervals). 
For this reason, confidence intervals are not included in Figures \ref{fig:censorKDE} or \ref{fig:violin}.

\subsubsection{High-throughput construction of models}

In order for models to be constructed in a reasonable amount of time, we used the University of Wisconsin – Madison's HTCondor framework \citep{Thain2005}. 
HTCondor is an open-source high-throughput computing infrastructure that distributes work among many execute nodes running in parallel. 
We used HTCondor version 8.5.1 which allows for encryption of data transferred to and from worker nodes. 
Since a separate model is built for each ICD-9 code, pan-diagnostic machine learning can easily be partitioned into parallel tasks by assigning a separate machine to build a predictive model for each diagnosis code. 
 Our term ``high-throughput machine learning'' acknowledges our intellectual debt to the computing philosophy of HTCondor and the high-throughput computing environment provided by UW-Madison's Center for High-Throughput Computing (CHTC).

\subsection{Simulated Prospective Study}

\subsubsection{Simulated prospective study design}

We simulated a prospective study by randomly selecting a test cohort of patients and forming predictions across all diagnoses for these patients over a one-year period. 
We chose to perform our study during the calendar year of 2014 as this maximized the amount of usable training data for building predictive models. 
A test cohort was generated by randomly selecting 100,000 patients who had at least one contact with the RFRHS in the calendar year of 2013 and were not known to be deceased at the end of 2013. 
This was done to simulate how a health system would perform a prospective study, by first identifying a cohort of active patients to follow.

Two sets of ICD-9 code models were built using pan-diagnostic machine learning with a 1-month prediction window and a 6-month prediction window. 
These models were constructed in the same fashion as the models built for the pan-diagnostic machine learning experiments. 
Training data for each model was produced by randomly selecting a set of patients using 1:1 case-control matching, as well as adhering to the prerequisite diagnoses established by DDR. 
Additionally, no patients present in the testing cohort were a part of the training data. 
All training data following the calendar year of 2013 were truncated to prevent leakage of any information.

For each ICD-9 code model, a subset of the testing cohort was selected based on eligibility criteria, to avoid overly-optimistic accuracy estimates from predicting for patients for whom such predictions are trivial to make correctly. 
The eligibility criteria were determined by the case patients chosen for training the model and included the following: 1) the test patient must have all DDR prerequisite diagnoses, 2) the test patient must have been aged between the 1st and 99th percentile of ages of the training case patients, and 3) the test patient must have been of the same sex as the majority of the training case patients if there was a 99\% or greater proportion of one sex. 
This ensured that the results of our models were not artificially elevated by testing on patients who would not have been considered for this diagnosis in a healthcare setting (e.g. 
prostate cancer for women, or Alzheimer's disease for an adolescent). 
Each eligible test patient received a score from the model as their predicted value and their true value was indicated by entries on their record during the year of 2014. 

\subsubsection{Simulated prospective study analysis}

The results of the prospective study yielded, for each modeled ICD-9 code and the subset of the testing cohort eligible for the model, the risk scores as predicted by the model on this subset, and the corresponding ground truth indicating if a given patient received the predicted code during the calendar year of 2014. 
For each diagnosis code, we required that there be sufficient patients to construct both a 1-month and 6-month predictive model. 
For a given patient and ICD-9 code, the risk for that patient was calculated as the maximum of the risks predicted by the 1-month and 6-month models. 
In this way, we roughly predict risk in the 1-year window of our study. 
Like the analysis performed for the experiment shown in Figure \ref{fig:censorKDE}, we use KDE to visualize the distribution of model AUCs in the simulated prospective study. 
However, the models produced in the first experiment had a minimum of 500 case and 500 control patients each. 
That amount of data is large enough to produce very good estimates of the AUC for each model. 
In the prospective study, there were significantly less positive patients for each ICD-9 code, with the majority of models having less than 10 patients who received the code during 2014. 
Fewer patients, in combination with heavy class skew, led us to much less tight estimates of AUC for each model; hence for the prospective trial, we also show 95\% confidence intervals of AUC for each model. 
To compute these confidence intervals, we first calculated the standard deviation using the formula \citep{Bamber1975} in Equation \ref{eq:auc_sigma}, where $n_p$ and $n_n$ are the number of positive (case) and negative (control) patients respectively, and AUC represents the AUC of a particular model.

\begin{equation}
\label{eq:auc_sigma}
\scriptstyle
\sigma = \sqrt{\dfrac{AUC(1-AUC) + (n_p - 1)(P_{xxy} - AUC^2) + (n_n - 1)(P_{xyy} - AUC^2)}{n_p n_n}}
\end{equation}

\noindent\begin{minipage}{.5\linewidth}
\begin{equation*}
    P_{xxy} = \dfrac{AUC}{2 - AUC}
\end{equation*}
\end{minipage}
\noindent\begin{minipage}{.5\linewidth}
\begin{equation*}
    P_{xyy} = \dfrac{2 AUC^2}{1 + AUC}
\end{equation*}
\end{minipage}
\vspace{.5cm}

We then computed the upper and lower bounds of the two-sided 95\% confidence interval as AUC $\pm z_{.025} \sigma$ where $z_\alpha$ was calculated via the percent point function for the normal distribution N(0,1). 
As AUC can only take on values between 0 and 1, bounds were truncated at 0 and 1: computed lower bounds below 0 were set to 0, and computed upper bounds above 1 were set to 1. 
We then produced two additional KDE distributions of the lower and upper bound AUC values. 
Note that some of these confidence interval densities, in addition to the density of AUC scores, have a small mass outside of the 0,1 range due to kernel density estimation, specifically the use of a Gaussian kernel to estimate the densities. 
This confidence interval calculation requires that the number of positive and negative examples not be too small or else the AUC may not be normally distributed. 
For this reason, we required each diagnosis code to have a minimum of 30 case and 30 control patients in the study cohort to be included in the results of this work. 
We retained a total of 2,538 models that contributed to the results presented in Figure \ref{fig:propKDE}A, and 2,130 models that for Figure \ref{fig:propKDE}B.

Figures \ref{fig:propKDE}A and \ref{fig:propKDE}B differ based on the inclusion (A) or exclusion (B) of so-called repeat diagnoses. 
In Figure \ref{fig:propKDE}B we predict a diagnosis for a patient only if that patient has never had this diagnosis on their record. 
In this way we are performing a similar task to our original experiments presented in Figures \ref{fig:censorKDE} and \ref{fig:violin}. 
However, in Figure \ref{fig:propKDE}B we predict an outcome for a patient even if they have had that diagnosis before. 
There are many acute diseases for which an additional code corresponds to a unique event separate from previous entries; however, there are also many other diseases for which this is not true. 
Including only first diagnoses may be somewhat pessimistic for some diseases, but including repeat diagnoses for all diseases is certainly overly optimistic. 
We believe the true aggregate performance falls somewhere between the two.

\subsubsection{In-depth analysis of high impact diseases}

In Figure \ref{fig:deepDive} we present an in-depth analysis of the models produced for three high impact diseases. 
The results for each model included only patients who met the model's eligibility criteria and had no entries of the ICD-9 code prior to 2014. 
For each disease, we generated a modified Kaplan-Meier curve, a ROC-curve, and a PR-curve. 
The modified Kaplan-Meier curve details the relationship between time since prediction (start of 2014) and the fraction of patients yet to receive the diagnosis. 
The ROC-curve depicts the relationship between the false positive rate and the true positive rate as the threshold is varied. 
The PR-curve shows the relationship between precision (positive predictive value) and recall (sensitivity, or true positive rate) as the threshold is varied.

\subsection{Additional Details}

\subsubsection{Additional software tools used}

Our pan-diagnostic machine learning code also makes use of NumPy \citep{VanDerWalt2011} and Pandas \citep{McKinney2010}. 
Figures were constructed using matplotlib \citep{Hunter2007} and seaborn \citep{Waskom}.

\section{Results}

\subsection{Prediction quality depends on disease category and time window}

We demonstrate the efficacy of the pan-diagnostic machine learning pipeline on the tasks of predicting diagnosis risks at the seven different time windows. 
The prediction window of 1 month is achieved by truncating all training data following the date 1 month prior to the case patient's diagnosis; each control patient for a given ICD9 code belongs to a case-control pair and has his/her data truncated at the same date as does the case patient. 
 An analogous process is executed for the six other prediction time windows. 
We observed mean AUCs ranging from 0.803 $\pm$ 0.062, across the 3,586 1-month models, to 0.524 $\pm$ 0.028 across the 3,288 20-year models. 
Fewer models were constructed at the 20-year window than at the 1-month window. 
Because we required a minimum number of case-control pairs that all have data prior to the truncation window, some models had sufficient patients for the 1-month window but not the 20-year window. 
We use KDE to visualize and compare the distributions of model performances when predicting at the several time intervals (Figure \ref{fig:censorKDE}). 
Additionally, we investigate how the predictive accuracy varies across and within the fifteen different highest-level chapters, or diagnosis categories, of the ICD hierarchy (Figure \ref{fig:violin}), using the 1-month and 6-month prediction time windows.

We see in Figure \ref{fig:censorKDE} that the majority of diagnoses (with sufficient data) can be predicted 1-month in advance with an AUC of 0.8 or greater. 
We note an inverse relationship between the length of the prediction window and the quality of the model. 
This observation is likely owed to the decrease in number of patients available for long term predictions, the smaller amount of data these patients have as we necessarily have access to less of their records, and the importance of a patient's recent health state on their immediate future. 
Interestingly, there are some diagnoses which can predict with AUC at or above 0.7 even 20 years in advance. 
These diagnoses were largely ocular disorders and congenital anomalies.

\begin{figure}
    \centering
    \includegraphics[width=\linewidth]{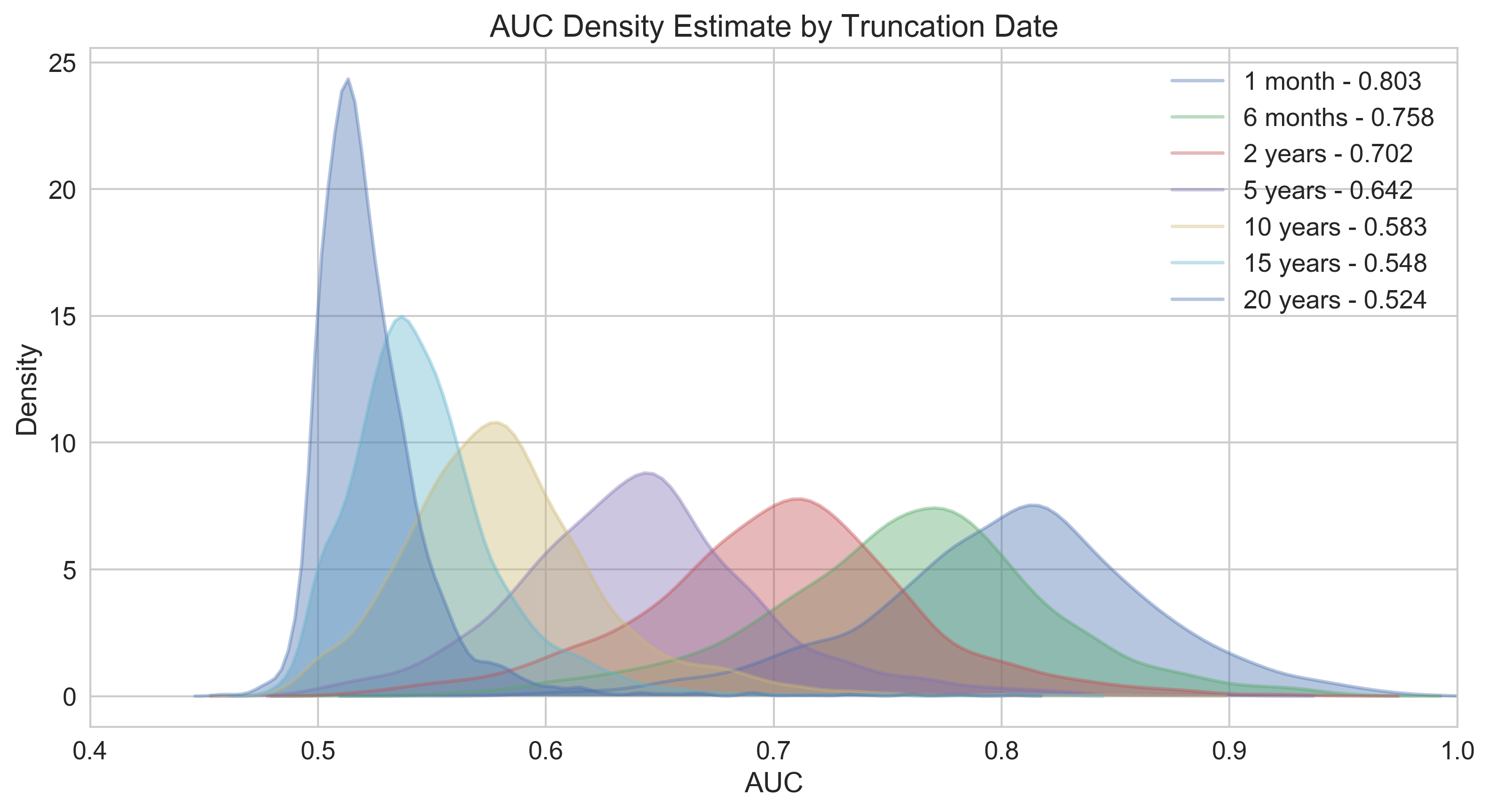}
    \caption[Comparison of Kernel Density Estimate (KDE) of AUC Distributions by Truncation Date.]{
            Comparison of Kernel Density Estimate (KDE) of AUC Distributions by Truncation Date.
            The shaded regions show the KDE distributions of AUC for the models built to predict ICD-9 codes at various time intervals ranging from 1 month to 20 years prior to first diagnosis. Note the significant performance increase attributed to an additional data leading up to diagnosis. 
            Additionally, we find it of interest that these distributions are approximately normal. 
            Some of the distributions, in particular those predicting far in advance, have a heavy right tail stretching well into AUCs of 0.7 and greater. 
            }
    \label{fig:censorKDE}
\end{figure}

\begin{figure}
    \centering
    \includegraphics[width=\linewidth]{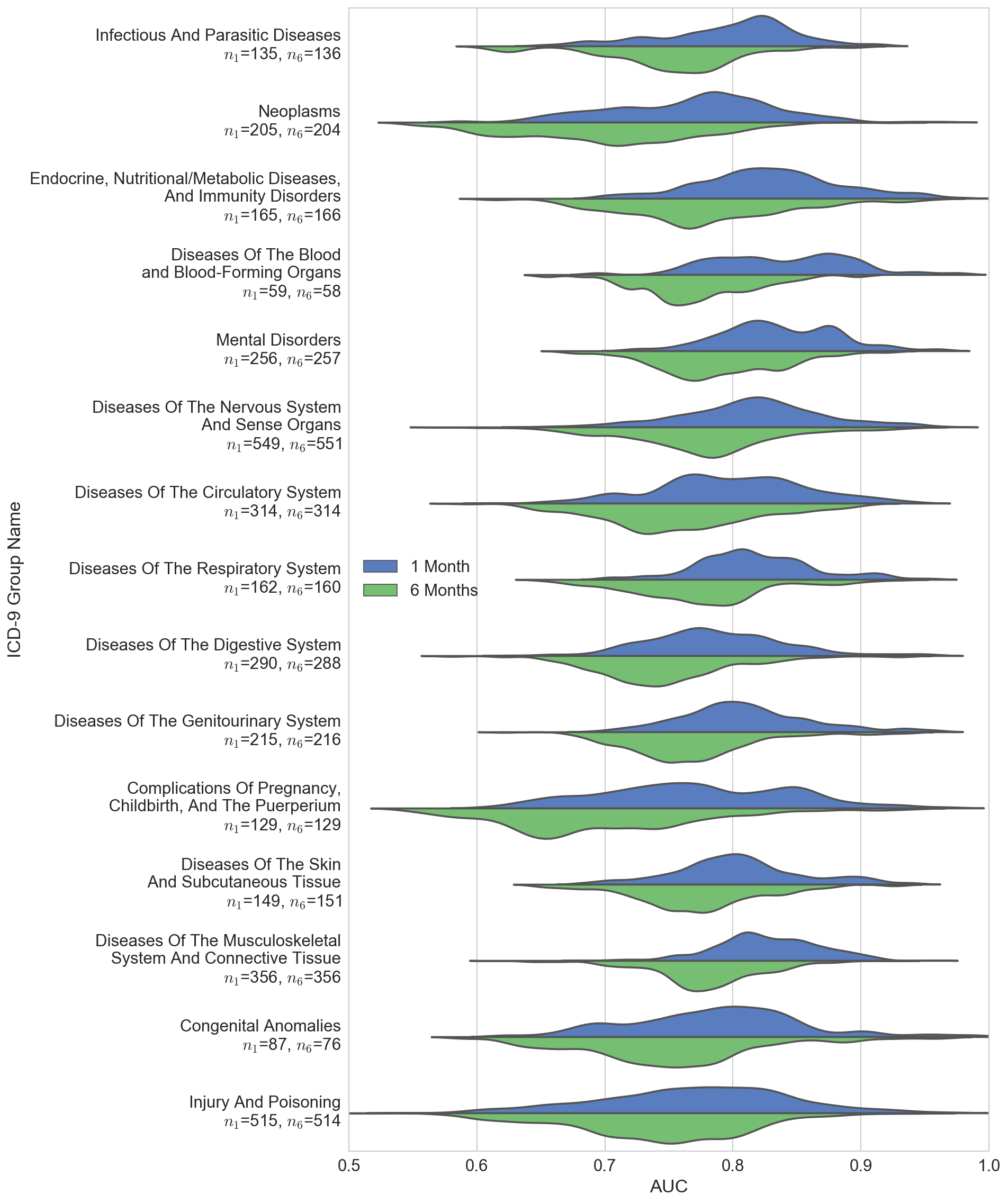}
    \caption[Split Violin Plot Comparing Kernel Density Estimates of AUC Distribution by Truncation Date and ICD-9 Diagnosis Group.]{(Continued on the following page.)}
    \label{fig:violin}
\end{figure}

\begin{figure}[t]
    \contcaption{
            Split Violin Plot Comparing Kernel Density Estimates of AUC Distribution by Truncation Date and ICD-9 Diagnosis Group.
            Across each diagnosis group a comparison of KDE distributions of AUC is shown for truncation dates of 1 and 6 months in the blue and green curves respectively. 
            The number of codes predicted at 1 month and 6 months are listed respectively. 
            The largest decrease in performance is among pregnancy complications where the mean AUC drops from $0.772 \pm 0.075$ at 1 month to $0.697 \pm 0.077$ at 6 months. 
            Additionally, note the bi- and sometimes tri-modal nature of distributions, suggesting that many of these categories may have interesting clusters within them.
            }
\end{figure}

	Encapsulated in Figure \ref{fig:censorKDE} are the AUC scores of some 24,462 models across the 7 different truncation windows. 
While this figure provides a high-level understanding of the general trends of how predictive efficacy varies with time, and while Figure \ref{fig:violin} delves deeper in presenting how individual diagnostic chapters vary from one another, there is a tremendous opportunity for exploring the results of this work in greater detail. 
For example, each random forest model has feature importance \citep{Breiman2001} values corresponding to an estimate of how valuable a particular health event is in predicting a diagnosis. 
These variable importance values might be used to better understand both individual diseases and how different diseases relate to one another. 
Furthermore, a more fine-tuned approach could be taken to the analysis by exploring a subset of diagnoses or how the prediction of a single diagnosis code changes over time. 
Due to the wealth of opportunities for additional research, we publish alongside this work a dataset containing the AUC and feature importance values for all 24,462 models contributing to Figure \ref{fig:censorKDE}. 
We believe that this dataset will be both a source for additional research and a baseline of comparison for other prediction strategies in future research. 
We note that because the random forest algorithm is stochastic that the feature importance values are as well (that is they can vary between runs). 
However, with a sufficient number of trees, such variance is minimal. 
Nevertheless, it is critical to inspect the features returned as a sanity check on the model, and thus in Table \ref{tab:feat_imp} we present top feature importances for three models predicting common health conditions: acute myocardial infarction, lung cancer, and influenza. 
We find that these models are largely using reasonable features matching known risk factors for their respective conditions.

\begin{table}[]
\small
\caption[Deep dive feature importance values for acute myocardial infarction, lung cancer, and influenza.]{
        Feature importance values for three common diseases. 
        Feature importance values provide a window into which variables are most important in differentiating between case and control patients in a given model. 
        However, this view is limited in that feature importance values cannot capture interactions among features and the feature importance values are strictly positive and thus cannot be used to determine if presence of the feature increases or decreases risk. 
        We present the top 10 unique features for each model, i.e., if a feature was useful in more than one time window, it was collapsed into a single feature for clarity. 
        Note that while we controlled for age, sex, and date of birth, these features are consistently top predictors and they still have value through interactions which other features. 
        Features seen in the acute myocardial infarction model match known risk factors. 
        Of interest is the ``Procedure Cytopathology, Pap Smear'' feature in lung cancer; we believe this feature is useful as it is a proxy for sex and also is a procedure only performed at specific ages. 
        Similarly, we believe the ICD-9 367, seen in two of the models, may act as a useful proxy for age as eyesight worsens with age. 
        The influenza model shows a mix of respiratory related features and features suggesting regular contact with the healthcare system (office visits and blood collection). 
        We believe that features suggesting greater contact with the healthcare system may be used to differentiate sicker patients from healthier patients, which could indicate a greater risk for influenza.
}
\label{tab:feat_imp}
\vspace{.3cm}
\begin{tabular}{p{1cm}|p{4cm}|p{4cm}|p{4cm}}
Rank & 410.4 Acute Myocardial Infarction (AUC 0.703)         & 162.9 Lung Cancer (AUC 0.728)                 & 487.1 Influenza (AUC 0.836)                         \\
\midrule
1    & Date of Birth                                         & Date of Birth                                 & Date of Birth                                       \\
2    & Age                                                   & Age                                           & Age                                                 \\
3    & ICD-9 410.9 Unspecified Essential Hypertension        & ICD-9 305.1 Nondependent Tobacco Use Disorder & ICD-9 465 Acute Upper Respiratory Infections        \\
4    & ICD-9 414 Ischemic Heart Disease                      & ICD-9 305 Nondependent Abuse of Drugs         & Procedure Office Visit                              \\
5    & ICD-9 305.1 Nondependent Tobacco Use Disorder         & ICD9 496 Chronic Airway Obstruction           & ICD-9 V72 Special Investigations and Examinations   \\
6    & Sex                                                   & ICD-9 518 Other Diseases of Lung              & ICD-9 780 General Symptoms                          \\
7    & ICD-9 305 Nondependent Abuse of Drugs                 & Sex                                           & Procedure Routine Venipuncture Collection           \\
8    & ICD-9 786 Respiratory System and Other Chest Symptoms & ICD-9 786.6 Swelling, Mass, or Lump in Chest  & ICD-9 462 Acute Pharyngitis                         \\
9    & ICD-9 367 Disorders of Refraction and Accommodation   & ICD-9 518.8 Other Diseases of Lung            & ICD-9 490 Bronchitis                                \\
10   & ICD-9 786.5 Chest Pain                                & Procedure Cytopathology, Pap Smear            & ICD-9 367 Disorders of Refraction and Accommodation
\end{tabular}
\end{table}

\subsection{Translational Validation via Simulated Prospective Study}

In addition to evaluation by cross-validation in our case-control samples, to better estimate the accuracy of pan-diagnostic machine learning if it were employed in a clinical setting, we performed a simulated prospective study. 
The goal of this experiment was to gain insight into the potential efficacy of the models if used in a live healthcare setting, by simulating the translation of our pan-diagnostic risk prediction system.

In our initial pan-diagnostic machine learning experiments we intentionally use a 1-to-1 case-control matching scheme. 
Because the lifetime incidence rates for diseases vary greatly from disease to disease, a 1-to-1 ratio can skew certain evaluation metrics such as precision and recall. 
Therefore, we utilize a truly random sample in our simulated prospective study to demonstrate results without case-control matching. 
Our prospective study ran during the calendar year of 2014 and followed 100,000 randomly selected patients who had visited the RFRHS at least once during 2013 and were alive as of the study start (see Simulated prospective study design in Methods and Materials).

\begin{figure}
    \centering
    \includegraphics[width=\linewidth]{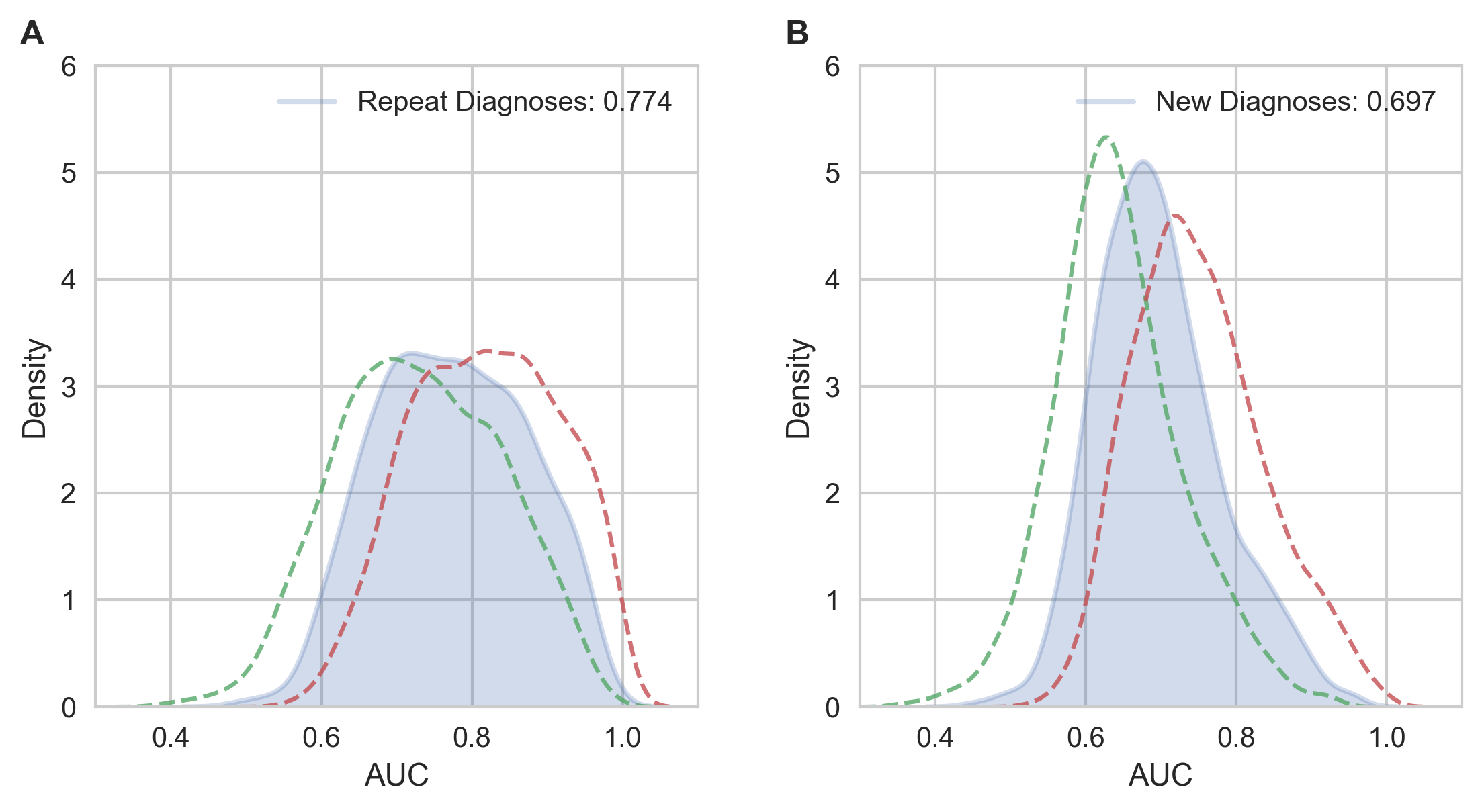}
    \caption[Comparison of KDEs of AUC Distributions by Minimum Number of Positive Patients and Inclusion of Repeat Diagnosis.]{
            Comparison of KDEs of AUC Distributions by Minimum Number of Positive Patients and Inclusion of Repeat Diagnosis. 
            In Figures \textbf{A} and \textbf{B}, the blue region represents the KDE distribution of AUC scores for models in the simulated prospective study; the green and red dotted lines represent the upper and lower bounds of the 95\% CI respectively. 
            \textbf{A}, 2,538 models with mean AUC 0.774, 95\% CI [0.730, 0.819]; including repeat diagnoses. 
            \textbf{B}, 2,130 models with mean AUC 0.697, 95\% CI [0.643, 0.751]; including only first diagnoses. 
            While including test patients who previously received the diagnosis provides a more optimistic distribution of AUC density, a conservative estimate of model efficacy would only include evaluation on new diagnoses.
            }
    \label{fig:propKDE}
\end{figure}

The prospective study assessed the predictive quality under two different paradigms, given the complexities in disease mechanisms. 
These paradigms addressed first diagnoses vs. 
repeat diagnoses. 
Some illnesses may be coded multiple times over a patient's life, but each diagnosis may be independent of one another (e.g., influenza), while other, chronic diseases may be coded multiple times as part of disease treatment (e.g., diabetes). 
We see higher performance when including repeat diagnoses as compared to the first entry of a code, illustrating the higher difficulty of predicting risks for first-time diagnoses. 
Figure \ref{fig:propKDE} shows a detailed view of the distribution of model efficacies when including or excluding repeat diagnoses. 
Only models corresponding to diagnosis codes with at least 30 case and 30 control patients in the 100,000-patient cohort were included in the prospective study evaluation. 
This minimum patient requirement was put in place as our evaluation metric, AUC, becomes unstable when one class has a small number of samples. 
For the task of predicting new diagnoses, we find that the prospective study shows a slight decrease in performance as compared to the models showcased in our initial experiments shown in Figure \ref{fig:censorKDE}. 
In our initial experiments with 1:1 case-control matching we achieved a mean AUC of 0.702 predicting diagnoses 2 years in advance. 
In the simulated prospective study, we see a mean AUC of 0.697 predicting over the course of a year. 
One potential explanation for this decreased AUC is the inclusion of patients who did not return during the study year in the evaluation data set. 
We included these patients in the evaluation set as their exclusion would provide a biased view of the results. 

\subsection{Limitations of our Experiments}

While receiver operating characteristic (ROC) curves are arguably the most common summary of predictive ability in both machine learning and medicine, a weakness (and a strength) is that they are not sensitive to disease incidence, or the natural ``class skew'' of cases to controls. 
In many domains with strong class skew, such as information retrieval and web search, precision-recall (PR) curves are used instead. 
Nevertheless, PR curves are less appropriate for comparing the results of many disease risk models at once because their different class skews can artificially make one disease look easier to predict than another. 
But for any one disease at a time, they can provide useful insights, so we explore PR curves and other metrics further in the next subsection, in the context of a few specific diseases of interest. 
 
Another limitation of even our prospective trial is that we do not compare our predictive models with current clinical practice, to see if any of them would currently improve clinical care. 
 A true, randomized prospective trial could provide such a comparison but is beyond the scope of our present work; our present work simply shows how well all coded diagnoses can be predicted across different lengths of time into the future. 
A few such predictive models have in fact already been translated into the clinic by others, such as the Framingham model for cardiovascular risk \citep{Dawber1951}, the Gail model for breast cancer risk \citep{Gail1989}, and models used in the emergency room such as Charlson comorbidity index \citep{Charlson1987}. 
These models were constructed by human-selected features and logistic regression, rather than the full EHR and random forests. 
While the latter approach often yields slightly more accurate models than the former approach \citep{Weiss2012,lantz2016machine}, models such as the Framingham risk score and the Gail model include identified input features and features found in text, and also use such data for improved phenotyping, while our modeling approach used only de-identified coded data for privacy reasons \citep{Gail1989,Dawber1951}. 
Before a clinical trial and translation, each of our models could almost certainly benefit significantly from being retrained with the inclusion of such data. 

\subsection{A Closer Look at Some Specific Diagnosis Models}

\begin{figure}
    \centering
    \includegraphics[width=\linewidth]{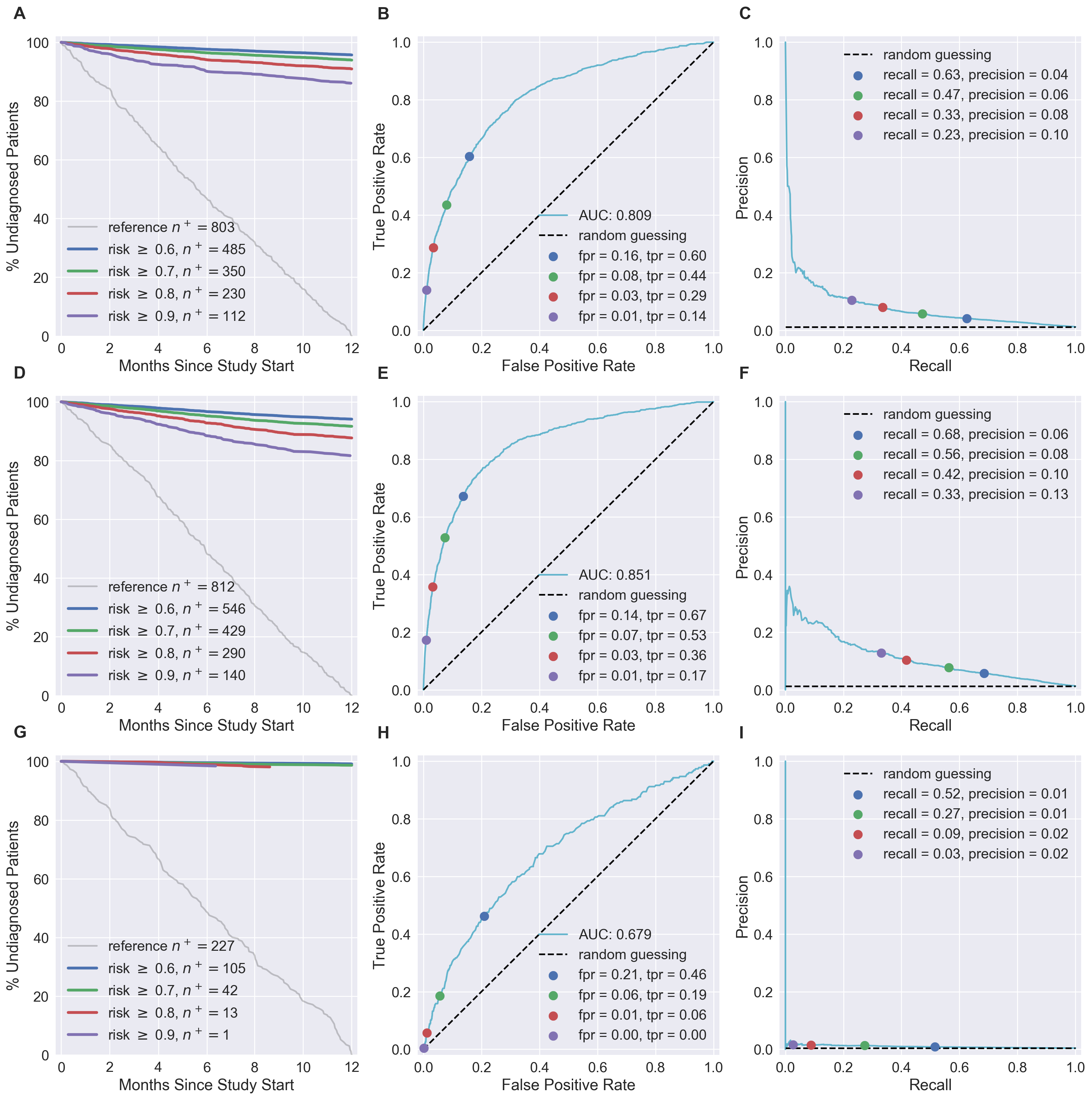}
    \caption[In-Depth Analysis of High Impact Diseases.]{
            In-Depth Analysis of High Impact Diseases. 
            Figures \textbf{A-C}, \textbf{D-F}, and \textbf{G-I} analyze Type II Diabetes (ICD-9 250.00), Chronic Kidney Disease (CKD) (ICD-9 585), and Acute MI (ICD-9 410) respectively. 
            In all figures, blue, green, red, and purple dots correspond to the prediction thresholds of 0.6, 0.7, 0.8, and 0.9 respectively. 
            \textbf{A}, Kaplan-Meier Curve for Type II Diabetes depicting the relationship between months since prediction and percent of predicted positives that had not received an entry of ICD-9 250.00 in the study year. 
            \textbf{B}, ROC-Curve for Type II Diabetes with overlaid operating points corresponding to risk thresholds identified in \textbf{A}. 
            \textbf{C}, PR-Curve for Type II Diabetes with overlaid operating points corresponding to risk thresholds identified in \textbf{A}. 
            \textbf{D}, Kaplan-Meier Curve for CKD. 
            \textbf{E}, ROC-Curve for CKD. 
            \textbf{F}, PR-Curve for CKD, note the poor performance in comparison to \textbf{E}. 
            \textbf{G}, Kaplan-Meier Curve for Acute MI. 
            \textbf{H}, ROC-Curve for Acute MI. 
            \textbf{I}, PR-Curve for Acute MI.
            }
    \label{fig:deepDive}
\end{figure}

In addition to the coarse analysis of our simulated prospective study, we present a deeper analysis of predicting the initial entry of three high impact diagnoses: type II diabetes \citep{Zimmet2001}, chronic kidney disease (CKD) \citep{Levey2005}, and acute myocardial infarction (MI) \citep{Tunstall-Pedoe1994} in Figure \ref{fig:deepDive}. 
Most notable from this analysis is the often tenuous relationship between ROC curve performance and PR curve performance. 
In learning tasks with high class skew (e.g., relatively rare diagnoses) ROC curves, which are skew independent, can give a much more optimistic picture of the results than do PR curves, which are skew dependent. 
We would like to aggregate PR areas as we did ROC areas in Fig 1, but because each ICD-9 code has a unique skew, aggregating PR areas across diagnoses is inappropriate \citep{Boyd2012}; therefore, instead we use these three individual PR curves to illustrate the limitations of our models despite their high overall distribution of ROC areas.

\section{Discussion}

This research introduces the concept of high-throughput machine learning and demonstrates its successful application to the task of predicting patient diagnosis risks, which we call pan-diagnostic machine learning. 
Additionally, we demonstrate an automated procedure for translation of these models into practice wherein appropriate subpopulations are automatically identified for each model. 
We find that many coded diagnoses can be predicted with reasonable performance—AUC greater than 0.8—at least one month prior to diagnosis as shown in Figure \ref{fig:censorKDE}. 
As we attempt to predict diagnosis risks further in advance, we become less confident in the quality of our predictions. 
We do, however, note that even 20 years prior to the first diagnosis, we can still predict with some efficacy the first entry of diagnosis codes categorized as congenital disorders (ICD-9 320-389) and diseases of the nervous system and sense organs (ICD-9 740-759). 
Amongst the 43 models predicting congenital anomalies, 10 were above 0.6 AUC (23\%) at this timeframe; amongst the 506 models predicting diseases of the nervous system and sense organs, 30 were above 0.6 AUC (5.6\%). 
Many of the highest performing models, in both congenital and nervous categories, were related to ocular disorders.

Of additional interest are the differences in predictive performance amongst models when stratified by the disease categories of the ICD-9 hierarchy as presented in Figure \ref{fig:violin}. 
The multi-modal distributions of some of these categories suggest that some of these categories can be further partitioned by predictive quality, such as those models predicting mental disorders or pregnancy complications at the 1-month window. 
We believe there are additional insights to gain from further analysis of these diagnoses. 
With a simulated prospective study, we estimate the translational efficacy of a pan-diagnostic machine learning pipeline in the healthcare setting. 
We note that the performances of these models under this evaluation methodology are slightly pessimistic as compared to the performances achieved in a 10-fold cross validation. 

We introduce a new medical dataset consisting of the AUC scores and feature importance values of the 24,482 predictive models belonging to Figure \ref{fig:censorKDE}. 
We believe that this dataset will be of interest both as a baseline for comparison for future work that predicts diagnoses and as a data source that can be mined for relationships amongst diagnoses and health record events. 
Many of our open questions such as which diagnoses cluster together on predictive quality and how predictive efficacy changes over time can be investigated with this dataset. 

The limitations of this research fall largely into two categories: those that are opportunities for improvements to various steps in our high-throughput machine learning pipeline, and those that arise from the aggregation of unique prediction tasks both in model building and evaluation. 
As the focus of this work is on the introduction of high-throughput machine learning and generating a baseline for the performance of predicting diagnosis code risks, we believe there are some areas of our pipeline that, while functional, could be improved to produce even higher quality models. 
For example, in place of rule-of-n for case-control definition, more advanced forms of electronic phenotyping \citep{Peissig2014} could be used. 
Future work could additionally incorporate existing disease ontologies to group together ICD-9 codes. 
Doing so would both provide computational benefits and aggregate codes into disease-specific models. 
Because our pipeline is flexible and modular, these enhancements could be easily created and incorporated into future research. 
Furthermore, it would be interesting to compare the results of this work to a multi-label learning task wherein a single model outputs risks for all diagnoses. 
This multi-label approach would allow us to leverage differences between diagnoses that co-occur versus those that occur independently. 
We additionally note that some limitations arise when creating general rules for the construction and evaluation of thousands of models. 
For example, by picking a single prediction time window, e.g. 
6-months, there are some diagnoses for which we are losing valuable data and others for which we would have preferred an even more greatly truncated window. 
For example, truncating by 6-months may be too strict for the prediction of an acute disease such as influenza, but may be too short of a window in the case of predicting a disease whose symptoms mount gradually leading to diagnosis, such as Parkinson's disease. 
While we present a variety of truncation windows between 1-month and 20-years for this reason, we do not have a data-driven method to automatically determine the appropriate truncation window for a given diagnosis. 
A final limitation of this work is the difficulty in analyzing the quality of any particular model relative to the current quality achieved at a clinic. 
Because we do not have baseline AUC values for how well healthcare providers can predict these diagnoses, it is impossible to make claims on the impact of any particular model. 
It is possible that a model with an AUC of 0.6 for one disease would be of more value than a model with an AUC of 0.8 for another disease if healthcare providers currently cannot predict the first disease at all, but can perfectly predict the second. 
This limitation is not unique to our work and it emphasizes the importance of clinical trials when considering the translation of any decision support tool into the clinic. 

In this paper, we demonstrate that a single system can predict risks for thousands of different coded diagnoses. 
This work makes available for public use the experimental results of our predictive models and the software were created to perform pan-diagnostic machine learning. 
Perhaps the most important contribution of this study is to provide an initial baseline for how accurately the wide range of disease phenotypes can be predicted from EHR data. 
We believe this is a step toward a much broader incorporation of machine learning-based prediction into clinical care.



\bibliographystyle{mcbride}

\newpage

\appendix
\section*{Appendix A}

\begin{table}[h!]
	\caption{Parameter values used to construct RandomForestClassifier}
	\label{tab:rf_params}
	\begin{tabular}{ll}
		Parameter                   & Value  \\
		\midrule
		n\_estimators               & 500    \\
		criterion                   & `gini' \\
		max\_depth                  & None   \\
		min\_samples\_split         & 2      \\
		min\_samples\_leaf          & 1      \\
		min\_weight\_fraction\_leaf & 0.0    \\
		max\_features               & 0.1    \\
		max\_leaf\_nodes            & None   \\
		bootstrap                   & True   \\
		oob\_score                  & False  \\
		n\_jobs                     & 1      \\
		random\_state               & None   \\
		verbose                     & 0      \\
		warm\_start                 & False  \\
		class\_weight               & None  
	\end{tabular}
\end{table}

\begin{table}[h!]
	\caption{Parameter values input to the function seaborn.kdepolot to construct Figures \ref{fig:censorKDE} and \ref{fig:propKDE}}
	\label{tab:kde_params}
	\begin{tabular}{llll}
		Parameter     & Value (Fig. \ref{fig:censorKDE}) & Value (Fig. \ref{fig:propKDE}) & Value (Fig \ref{fig:propKDE} CI) \\
		\midrule
		shade         & True           & True           & False            \\
		vertical      & False          & False          & False            \\
		kernel        & `gau'          & `gau'          & `gau'            \\
		bw            & 0.008          & `scott'        & `scott'          \\
		gridsize      & 100            & 100            & 100              \\
		clip          & None           & None           & None             \\
		legend        & True           & False          & False            \\
		cumulative    & False          & False          & False            \\
		shade\_lowest & True           & True           & True            
	\end{tabular}
\end{table}

\begin{table}[]
	\caption{Parameter values input to the function seaborn.violinplot to construct Figure \ref{fig:violin}}
	\label{tab:violin_params}
	\begin{tabular}{ll}
		Parameter  & Value (Fig. \ref{fig:violin}) \\
		hue\_order & None                                           \\
		bw         & 0.2                                            \\
		cut        & 2                                              \\
		scale      & `area'                                         \\
		scale\_hue & True                                           \\
		gridsize   & 100                                            \\
		width      & 0.8                                            \\
		inner      & None                                           \\
		split      & True                                           \\
		orient     & 'h'                                            \\
		linewidth  & None                                           \\
		color      & None                                           \\
		palette    & `muted'                                        \\
		saturation & 0.75                                          
	\end{tabular}
\end{table}

\begin{table}[t]
	\caption{ICD-9 Diagnosis codes not allowed to be used as a prerequitie diagnosis in DDR.}
	\label{tab:icd9_excl}
	\begin{tabular}{p{1cm}p{12cm}}
		Code & Description                                                                  \\
		\midrule
		89   & Diagnostic Interview, Consultation, and Evaluation                           \\
		V20  & Health supervision of infant or child                                        \\
		V24  & Postpartum care and examination                                              \\
		V25  & Encounter for contraceptive management                                       \\
		V28  & Encounter for anteatal screening of mother                                   \\
		V29  & Observation and evaluation of newborns for suspected conditions not found    \\
		V51  & Aftercare involving the use of plastic surgery                               \\
		V56  & Encounter for dialysis and dialysis catheter care                            \\
		V58  & Encounter for other and unspecified procedures and aftercare                 \\
		V60  & Housing household and economic circustances                                  \\
		V61  & Other family circumstances                                                   \\
		V62  & Other psychosocial circumstances                                             \\
		V63  & Unavailability of other medical facilities for care                          \\
		V64  & Persons encountering health services for specific procedures not carried out \\
		V65  & Other persons seeking consultation                                           \\
		V66  & Convalescence and palliative care                                            \\
		V67  & Follow-up examination                                                        \\
		V68  & Encounters for administrative purposes                                       \\
		V69  & Problems related to lifestyle                                                \\
		V70  & General medical examination                                                  \\
		V71  & Observation and evaluation for suspected conditions not found                \\
		V72  & Special investigations and examinations                                      \\
		V73  & Special screening examination for viral and chlamydial diseases              \\
		V74  & Special screening examination for bacterial and spirochetal diseases         \\
		V75  & Special screening examination for other infectious diseases                  \\
		V76  & Special screening for malignant neoplasms                                    \\
		V77  & Special screening for endocrine nutritional metabolic and immunity disorders \\
		V78  & Special screening for disorders of blood and blood-forming organs            \\
		V79  & Special screening for mental disorders and developmental handicaps           \\
		V80  & Special screening for neurological eye and ear diseases                      \\
		V81  & Special screening for cardiovascular respiratory and genitourinary diseases  \\
		V82  & Special screening for other conditions                                      
	\end{tabular}
\end{table}

\end{document}